# Classical Interpretation to Relatively Stable Physical Statistical Distributions

Wei-Long She

Institute for Lasers & Spectroscopy, Sun Yat-Sen University, Guangzhou, China

Very recently we present a theory to discuss the nature of light and show that the quantization of light energy in vacuum can be derived directly from classical electromagnetic theory. In the theory a key concept of stability of statistical distribution and a variational procedure searching for those of relatively stable statistical distributions were introduced. This is a classical interpretation to the concept and method concerned. It would be helpful for avoiding some of possible obsession from the variational procedure and its final quantum result.

Very recently we present a theory [1] to show that the quantization of light energy in vacuum can be derived directly from classical electromagnetic theory through the consideration of statistics based on classical physics and reveal that the quantization of energy is an intrinsic property of light as a classical electromagnetic wave and has no need of being related to particles. Here is an interpretation to the theory, which would be helpful for understanding the concept and method concerned and avoiding some of obsession possible.

Suppose $\rho(q)$ be the density function of the generalized coordinate of 1-D harmonic oscillator (ODHO), $q$, to be found, then we can always introduce such a complex-valued function $\psi(q)$ mathematically that makes $|\psi(q)|^2 = \rho(q) \geq 0$. The $\psi(q)$ should then satisfy the following universal conditions:

$$\int_{-\infty}^{\infty} |\psi(q)|^2 \, dq = 1, \tag{1}$$

$$\lim_{|q| \to \infty} \psi(q) = 0, \tag{2}$$

$$\int_{-\infty}^{\infty} 1/2 \cdot \omega^2 q^2 |\psi(q)|^2 \, dq \leq E < \infty. \tag{3}$$

These are understandable ones in classical physics: Eq.(1) is the general condition for any density function; Eq.(2) ensures the amplitude of each ODHO being finite; and Eq.(3) is the finiteness condition of the expectation value, $E$, of the ODHO's energy. One can see that those meeting these conditions make up of the whole of the function candidates used to construct $\rho(q)$ and the number of them is infinite. To search for those of relatively stable statistical distributions of $q$, in Ref.[1] we changed Eqs.(1)-(3) into a set of equivalent "equilibrium conditions"(EECs)[1] (see below); introduced a correlative functional and then performed the variation. To make the concept and method mentioned easily understood and show they to be of the classical, let us draw an analogy as follows:

$$1/2\int_{-\infty}^{\infty}|\frac{d\psi(q)}{dq}|^2\,dq = 1/2\int_{-\infty}^{\infty}\omega^2 L^2\,|F(L)|^2\,dL,$$

$$1/2\int_{-\infty}^{\infty}|\frac{dF(L)}{dL}|^2\,dL = 1/2\int_{-\infty}^{\infty}\omega^2 q^2\,|\psi(q)|^2\,dq < \infty,$$

$$\int_{-\infty}^{\infty}|F(L)|^2\,dL = 1,$$

$$\int_{-\infty}^{\infty}|\psi(q)|^2\,dq = 1,$$

$$\lim_{|q|\to\infty}\psi(q) = 0,$$

$$\lim_{|q|\to\infty}\int_{-\infty}^{\infty}F(L)\exp(-i\omega qL)dL = 0.$$

The law $\psi(q)$ and $F(L)$ obey (EECs, see Ref. [1])

The really physical statistical distribution of $q$ cannot be just $|\psi(q)|^2$ [Fig.1(a)] coming from the rigid solution of EECs [1]. It will encounter some unpredictable but ineluctable tiny perturbations, $\delta\psi$, independent of the solutions of EECs, i.e., it should be $|\psi(q)+\delta\psi|^2$ but unpredictable [Fig.1(b)].

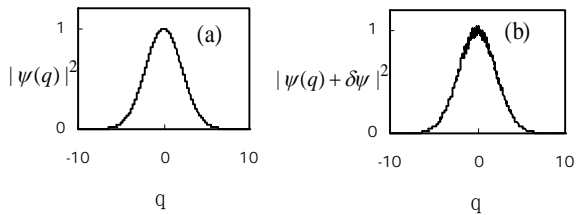

Fig.1 the comparison between an ideal statistical distribution of $q$ and a real one: (a) ideal; (b) real.

All the rigid solutions of EECs [about $\psi(q)$] can be divided into two types. One is scarcely different from $\psi(q)+\delta\psi$ in the sense of fitting EECs, and another is not. The first one resists the perturbations and is relatively stable while the second is too sensitive to the perturbations, deviating from EECs fleetly when the perturbations appear. So it cannot exist really in physics. Obviously, the relatively stable solutions of EECs should be those of optimized solutions of the following overdetermined equations (built for $\psi(q)+\delta\psi$ to fit EECs and $\delta\psi$ to take a series of its possible forms):

$$1/2\int_{-\infty}^{\infty}|\frac{d[\psi(q)+\delta\psi_i]}{dq}|^2\,dq - 1/2\int_{-\infty}^{\infty}\omega^2 L^2\,|F(L)+\delta F_i|^2\,dL = 0,$$

## Analogue

$$x^2 - y^2 = 0 \quad (4)$$

The law $x$ and $y$ obey (assumed)

The really physical quantities cannot be just the $x$ and $y$ coming from the rigid solution of Eq.(4). They will encounter some unpredictable but ineluctable tiny perturbations, $\delta x$ and $\delta y$, independent of the solution of the equation, i.e., they should be $x+\delta x$ and $y+\delta y$ but unpredictable. All the rigid solutions of Eq.(4) can be divided into two types. One is scarcely different from ($x+\delta x$, $y+\delta y$) in the sense of fitting Eq.(4), and another is not (Fig.2).

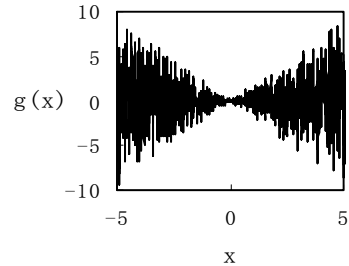

Fig.2 the stability of the solutions of equation $x^2 - y^2 = 0$. $g(x) = (x+\delta x)^2 - (y+\delta y)^2$ ($y = x$), where $\delta x$ and $\delta y$ are two small random numbers; Similarly when $y = -x$.

The first one [i.e.(0,0)] resists the perturbations and is relatively stable while the second is too sensitive to the perturbations, deviating from the law Eq.(4) fleetly when the perturbations appear (Fig.2). So it

$1/2\int_{-\infty}^{\infty}|\frac{d[F(L)+\delta F_i]}{dL}|^2\,dL-1/2\int_{-\infty}^{\infty}\omega^2q^2|\psi(q)+\delta\psi_i|^2\,dq=0$,

$\int_{-\infty}^{\infty}|F(L)+\delta F_i|^2\,dL-1=0$,

$\int_{-\infty}^{\infty}|[\psi(q)+\delta\psi_i]|^2\,dq-1=0$,

$\lim_{|q|\to\infty}[[\psi(q)+\delta\psi_i]]=0$,

$\lim_{|q|\to\infty}\int_{-\infty}^{\infty}[F(L)+\delta F_i]\exp(-i\omega qL)dL=0$,

(i = 1,2,3,..., n; n >> 2)

where $\delta\psi_1,\delta\psi_2,...\delta\psi_n$ stand for the possible forms of $\delta\psi$ and $\delta F_1,\delta F_2,...\delta F_n$ stand for those of $\delta F$ introduced on the consideration of symmetry [the effects of $\delta\psi$ on $\psi(q)$ and $\delta F$ on $F(L)$ are equivalent for fitting EECs], which is the tiny perturbations on $F(L)$ independent of the solutions of EECs and therefore independent of $\delta\psi$.

**The optimized solutions mentioned can be obtained by introducing functional (with two unknown functions)**

$I(\psi(q),F(L))=1/2\int_{-\infty}^{\infty}|\frac{d\psi(q)}{dq}|^2\,dq-1/2\int_{-\infty}^{\infty}\omega^2L^2|F(L)|^2\,dL$,

$1/2\int_{-\infty}^{\infty}|\frac{dF(L)}{dL}|^2\,dL-1/2\int_{-\infty}^{\infty}\omega^2q^2|\psi(q)|^2\,dq=0$,

$\int_{-\infty}^{\infty}|F(L)|^2\,dL-1=0$,

$\int_{-\infty}^{\infty}|\psi(q)|^2\,dq-1=0$,

$\lim_{|q|\to\infty}\psi(q)=0$,

$\lim_{|q|\to\infty}\int_{-\infty}^{\infty}F(L)\exp(-i\omega qL)dL=0$

and performing $\delta I(\psi(q),F(L))=0$ as did Ref.[1], which are just those satisfying the following equations:

$-1/2\cdot d^2/dq^2\cdot\psi(q)+(-\lambda_1)1/2\cdot\omega^2q^2\psi(q)+\lambda_3\psi(q)=0$,

$-\lambda_1/2\cdot d^2/dL^2\cdot F(L)-1/2\cdot\omega^2L^2F(L)+\lambda_2F(L)=0$.

where $\lambda_i$ ($i$ =1, 2, 3) denote three Lagrange's multipliers.

cannot exist really in physics. Obviously, the relatively stable solution of Eq.(4) should be the optimized solution of the following overdetermined equations [built for ($x+\delta x$, $y+\delta y$) to fit Eq.(4) and $\delta x$ (or $\delta y$) to take a series of its possible values]:

$(x+\delta x_1)^2-(y+\delta y_1)^2=0$,

$(x+\delta x_2)^2-(y+\delta y_2)^2=0$,

……………………………

$(x+\delta x_n)^2-(y+\delta y_n)^2=0$,

(n >> 2)

where $\delta x_1,\delta x_2,...\delta x_n$ stand for the possible values of $\delta x$ and $\delta y_1,\delta y_2,...\delta y_n$ stand for those of $\delta y$, simulated by some of random numbers.

**The optimized solution mentioned can be obtained by introducing function (with two independent variables)**

$$f(x,y)=x^2-y^2$$

and performing

$\partial f/\partial x=0$,
$\partial f/\partial y=0$,

which is just $(0,0)$. Of course, the result can also be approached by using numerical method.

The idea described above comes from classical physics especially classical experimental physics.